\documentclass[reqno]{amsart}

\newcommand{\Rm}{\mathbb{R}}
\newcommand{\Cm}{\mathbb{C}}

\newcommand{\Zm}{\mathbb{Z}}

\newcommand{\be}{\begin{equation}}
\newcommand{\ee}{\end{equation}}
\newcommand{\ba}{\begin{equation}\begin{aligned}}
\newcommand{\ea}{\end{aligned}\end{equation}}
\newcommand{\va}{\varphi}

\newcommand{\pp}{\partial}

\newcommand{\rvv}[1]{\boldsymbol{\mathrm{#1}}}
\newcommand{\hvv}[1]{\boldsymbol{\hat{\mathrm{#1}}}}

\providecommand{\Keywords}[1]{\small\textbf{\textit{Keywords---}} #1}

\usepackage{amsmath}
\usepackage[foot]{amsaddr}
\usepackage{amsthm,amssymb,mathrsfs}
\usepackage{graphicx}
\usepackage{color}

\theoremstyle{remark}

\title{The depth of the banana and the impulse stripe illumination for diffuse optical tomography}

\author{Manabu Machida$^1$}
\author{Keita Osada$^2$}
\author{Keiichiro Kagawa$^3$}
\address{${^1}$Department of Informatics, Faculty of Engineering, Kindai University, Higashi-Hiroshima 739-2116, Japan}
\address{${^2}$Graduate School of Integrated Science and Technology, Shizuoka University, Hamamatsu 432-8011, Japan}
\address{${^3}$Research Institute of Electronics, Shizuoka University, Hamamatsu 432-8011, Japan}
\email{machida@hiro.kindai.ac.jp}

\date{\today}

\begin{document}

\begin{abstract}
The stripe illumination lies between the illumination in the spatial-frequency domain and the point illumination. Although the stripe illumination has a periodic structure as the illumination in the spatial-frequency domain, light from the stripe illumination can reach deep regions in biological tissue since it can be regarded as an array of point illuminations. For a pair of a source and a detector, the shape of light paths which connect the source and detector is called the banana shape. First, we investigate the depth of the banana. In the case of the zero boundary condition, we found that the depth of the center of the banana is about $0.2d_{\rm SD}$ for typical optical parameters, where $d_{\rm SD}$ is the distance between the source and detector on the boundary. In general, the depth depends on the absorption and diffusion coefficients, and the ratio of refractive indices on the boundary. Next, we perform diffuse optical tomography for the stripe illumination against forward data taken by Monte Carlo simulation. We consider an impulse illumination of the shape of a stripe. By this time-resolved measurement, the absorption coefficient of a target is reconstructed.
\end{abstract}

\maketitle


\section{Introduction}
\label{intro}

Conventionally in diffuse optical tomography, light has been illuminated and detected with optical fibers. Noncontact diffuse optical tomography can be achieved when a laser beam is sent to a sample to which no optical fiber is attached and the detected light is measured by a CCD or CMOS camera. As an alternative noncontact diffuse optical tomography, measurements in the spatial frequency domain has been proposed \cite{Cuccia-etal05} (See \cite{Cuccia-etal09,Angelo-etal19} and references therein).

When the measurement is performed in the spatial frequency domain, the penetration depth of near-infrared light can be controlled by the spatial frequency. Light with high spatial frequencies can reach shallow regions and is reflected back to the surface of the sample. In other words, spatially oscillating light cannot reach deep regions compared with a planar light which uniformly illuminates the surface or a pencil beam which is illuminated at a point on the surface.

In this paper, we consider diffuse optical tomography with the impulse stripe illumination. Since the stripe illumination can be regarded as the illumination with multiple spatial frequencies, on one hand, it is the measurement in the spatial frequency domain. On the other hand, the stripe illumination can be regarded as an array of point illuminations. Hence the stripe illumination acts as a bridge between measurements in the real spatial domain and spatial frequency domain. Moreover, we send an impulse of near-infrared light for the stripe illumination and consider the time-resolved measurement. See \cite{Takada-etal21}.

For the conventional near-infrared measurement of one point illumination and one point detection \cite{Hock-etal97,Quaresima-Ferrari19}, trajectories of the detected light in the sample is known to form a banana shape \cite{Weiss-etal89,Schotland93,Feng-etal95,Zhu-etal96,Boas-etal01,Sassaroli-etal14}. This photon-path structure also takes place for the stripe illumination. In this paper, we consider the depth of the banana. We found that the depth of the center of the banana is about $0.2$ of the source-detector distance for typical measurements for biological tissue.

The remainder of the paper is organized as follows. In Sec.~\ref{banana}, we consider the banana shape for diffuse light. In Sec.~\ref{psi}, the impulse stripe illumination is introduced. In Sec.~\ref{DOT}, we obtain tomographic images for the forward data computed by Monte Carlo simulation. Finally, concluding remarks are given in Sec.~\ref{concl}.

\section{Materials and methods}

\subsection{Banana}
\label{banana}

Let $\rvv{r}=(\rvv{\rho},z)$ be a vector in $\Rm^3$, where $\rvv{\rho}\in\Rm^2$ is a vector in the $x$-$y$ plane. Let $\Omega$ be the half-space ($z>0$):
\be
\Omega=\left\{\rvv{r}\in\Rm^3;\;-\infty<x<\infty,\;-\infty<y<\infty,\;0<z<\infty\right\}.
\ee
Let $\pp\Omega$ be the boundary of $\Omega$, i.e., the $x$-$y$ plane. Suppose that $\Omega$ is occupied by a medium in which near-infrared light propagates. The outside $\Rm^3\setminus\overline{\Omega}$ is air.

\begin{figure}[htbp]
\centering
\includegraphics[width=0.8\textwidth]{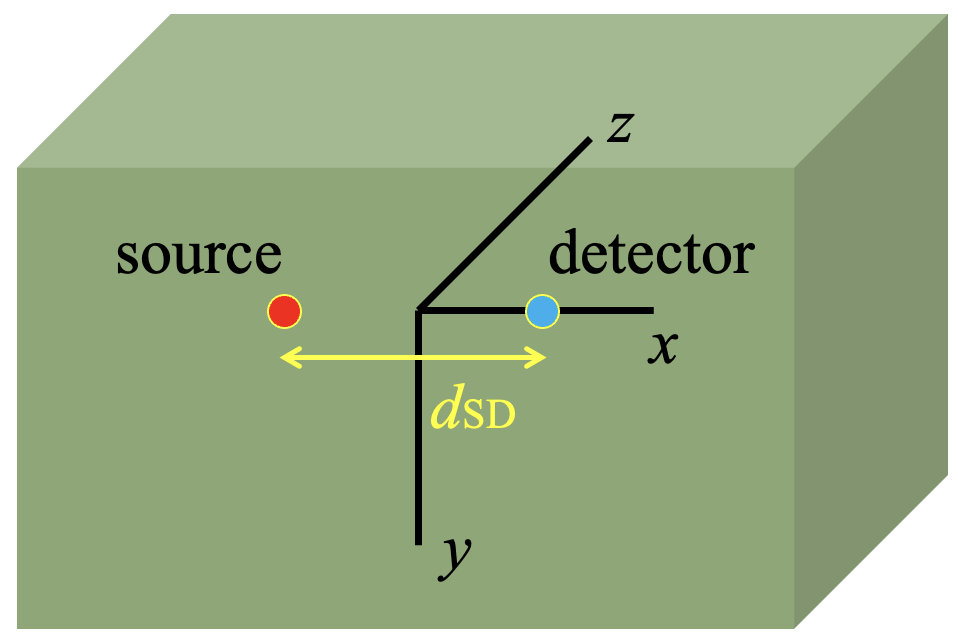}
\caption{
Light is illuminated at $(-d_{\rm SD}/2,0,0)$ and detected at $(d_{\rm SD}/2,0,0)$ on the boundary of the half space.
}
\label{fig_sd}
\end{figure}

Suppose that light is illuminated at $\rvv{r}_s\in\pp\Omega$ and detected at another point $\rvv{r}_d\in\pp\Omega$. Let $\rvv{r}_s={^t}(x_s,0,0^+)$ and $\rvv{r}_d={^t}(x_d,0,0)$ with $-x_s=x_d=d_{\rm SD}/2>0$, i.e, $d_{\rm SD}$ is the distance between the source and detector. See Fig.~\ref{fig_sd}. Let $u(\rvv{r})$ be the diffuse fluence rate at $\rvv{r}$. The light propagation is governed by the following the diffusion equation.
\be
\left\{\begin{aligned}
-D_0\Delta u+\bar{\mu}_au=\delta(\rvv{r}-\rvv{r}_s),
&\quad\rvv{r}\in\Omega,
\\
-D_0\frac{\pp}{\pp z}u+\frac{1}{\zeta}u=0,
&\quad\rvv{r}\in\pp\Omega,
\end{aligned}\right.
\label{de}
\ee
where $\bar{\mu}_a>0$ is the absorption coefficient and $D_0=1/(3\mu_s')$ is the diffusion coefficient with $\mu_s'$ the reduced scattering coefficient. We assume the diffuse surface reflection and give the constant $\zeta$ by \cite{Egan-Hilgeman79}
\be
\zeta=2\frac{1+R}{1-R},\quad
R=-1.4399\mathfrak{n}^{-2}+0.7099\mathfrak{n}^{-1}+0.6681+0.0636\mathfrak{n},
\ee
where $\mathfrak{n}$ is the refractive index of the medium.

To investigate the propagation of the detected light, we suppose that there is a point absorber at $\rvv{r}_0={^t}(x_0,0,z_0)$ and consider the diffusion equation given by
\be
\left\{\begin{aligned}
-D_0\Delta U+\mu_a(\rvv{r})U=\delta(\rvv{r}-\rvv{r}_s),
&\quad\rvv{r}\in\Omega,
\\
-D_0\frac{\pp}{\pp z}U+\frac{1}{\zeta}U=0,
&\quad\rvv{r}\in\pp\Omega.
\end{aligned}\right.
\label{de1}
\ee
Here,
\be
\mu_a(\rvv{r})=\bar{\mu}_a+\eta_0\delta(\rvv{r}-\rvv{r}_0),
\ee
where $\delta(\cdot)$ is the Dirac delta function and $\eta_0>0$ is a constant. The idea to detect the shape of the banana is similar to the idea used in the banana-shape regions theory \cite{Feng-etal95,Zhu-etal96,Sassaroli-etal14} in the sense that an absorber is assumed. However, we assume a point absorber instead of a sphere and will consider the depth of the banana instead of its shape. We assume that $\eta_0$ is small so that the Born approximation (below) holds.

The Green's function $G(\rvv{r},\rvv{r}')$ is introduced as
\begin{equation}
\left\{\begin{aligned}
-D_0\Delta G+\bar{\mu}_aG=\delta(\rvv{r}-\rvv{r}'),
&\quad\rvv{r}\in\Omega,
\\
-D_0\frac{\pp}{\pp z}G+\frac{1}{\zeta}G=0,
&\quad\rvv{r}\in\pp\Omega.
\end{aligned}\right.
\label{Agfunc:def}
\end{equation}
Thus, the following identity is derived.
\be
U(\rvv{r};\rvv{r}_0,\rvv{r}_s)=G(\rvv{r},\rvv{r}_s)-\eta_0G(\rvv{r},\rvv{r}_0)U(\rvv{r}_0;\rvv{r}_0,\rvv{r}_s)
\ee
With the Born approximation, we have $U\approx u_B$, where
\be
u_B(\rvv{r};\rvv{r}_0,\rvv{r}_s)=G(\rvv{r},\rvv{r}_s)-\eta_0 G(\rvv{r},\rvv{r}_0)G(\rvv{r}_0,\rvv{r}_s).
\ee
We have (see Appendix \ref{Agfunc})
\ba
G(\rvv{r}_0,\rvv{r}_s)
&=
\frac{z_e}{2\pi D_0}\int_0^{\infty}qJ_0(q|x_0-x_s|)
\frac{e^{-\lambda(q)z_0}}{1+\lambda(q)z_e}\,dq,
\\
G(\rvv{r}_0,\rvv{r}_d)
&=
\frac{z_e}{2\pi D_0}\int_0^{\infty}qJ_0(q|x_0-x_d|)
\frac{e^{-\lambda(q)z_0}}{1+\lambda(q)z_e}\,dq,
\ea
where
\be
\lambda(q)=\sqrt{\frac{\bar{\mu}_a}{D_0}+q^2},\quad z_e=\zeta D_0.
\ee
Here, $q\ge0$ is the magnitude of the Fourier vector $\rvv{q}$ (see (\ref{fourierG})). We note the reciprocal property $G(\rvv{r},\rvv{r}')=G(\rvv{r}',\rvv{r})$.

Since
\be
\frac{\pp}{\pp z_0}u_B(\rvv{r}_d;\rvv{r}_0,\rvv{r}_s)=
-\eta_0\left[\frac{\pp G(\rvv{r}_d,\rvv{r}_0)}{\pp z_0}G(\rvv{r}_0,\rvv{r}_s)+
G(\rvv{r}_d,\rvv{r}_0)\frac{\pp G(\rvv{r}_0,\rvv{r}_s)}{\pp z_0}\right],
\ee
we obtain
\ba
&
\frac{\pp}{\pp z_0}u_B(\rvv{r}_d;\rvv{r}_0,\rvv{r}_s)=
\eta_0\left(\frac{z_e}{2\pi D_0}\right)^2
\\
&\times
\Biggl[
\left(\int_0^{\infty}qJ_0(q|x_0-x_d|)\frac{\lambda(q)e^{-\lambda(q)z_0}}{1+\lambda(q)z_e}\,dq\right)
\left(
\int_0^{\infty}qJ_0(q|x_0-x_s|)\frac{e^{-\lambda(q)z_0}}{1+\lambda(q)z_e}\,dq\right)
\\
&+
\left(\int_0^{\infty}qJ_0(q|x_0-x_d|)\frac{e^{-\lambda(q)z_0}}{1+\lambda(q)z_e}\,dq\right)
\left(
\int_0^{\infty}qJ_0(q|x_0-x_s|)\frac{\lambda(q)e^{-\lambda(q)z_0}}{1+\lambda(q)z_e}\,dq\right)
\Biggr].
\ea

Let us investigate the position of the center of the banana shape by moving the point absorber. Recall that the solution of (\ref{de}) is indeed the Green's function. The reciprocal property of the Green's function implies that the banana is symmetric about $\rvv{r}_s$ and $\rvv{r}_d$. Hence, hereafter we set $x_0=0$.

If the point absorber is placed at the center of the banana, the detected light $u(\rvv{r}_d;\rvv{r}_0,\rvv{r}_s)$ should take the minimum value. In this case, we have $\frac{\pp}{\pp z_0}u_B(\rvv{r}_d;\rvv{r}_0,\rvv{r}_s)=0$. The condition implies
\be
\int_0^{\infty}qJ_0\left(\frac{qd_{\rm SD}}{2}\right)\frac{\lambda(q)e^{-\lambda(q)z_0}}{1+\lambda(q)z_e}\,dq=0.
\ee
By using $qdq=\lambda d\lambda$ and setting $x=\lambda d_{\rm SD}/2$, we have
\be
\int_{\sqrt{\frac{\bar{\mu}_a}{D_0}}}^{\infty}J_0\left(\frac{d_{\rm SD}}{2}\sqrt{\lambda^2-\frac{\bar{\mu}_a}{D_0}}\right)\frac{\lambda^2e^{-\lambda z_0}}{1+\lambda z_e}\,d\lambda
=\left(\frac{2}{d_{\rm SD}}\right)^3
\int_a^{\infty}J_0\left(\sqrt{x^2-a^2}\right)\frac{x^2e^{-wx}}{1+bx}\,dx
=0,
\ee
where
\be
a=\frac{d_{\rm SD}}{2}\sqrt{\frac{\bar{\mu}_a}{D_0}},\quad
b=\frac{2z_e}{d_{\rm SD}},\quad w=\frac{2z_0}{d_{\rm SD}}.
\ee
Thus the problem reduces to the problem of finding the zero of the following function $\Lambda(w)$:
\be
\Lambda(w;a,b)=\int_a^{\infty}J_0\left(\sqrt{x^2-a^2}\right)\frac{x^2e^{-wx}}{1+bx}\,dx.
\label{Lambdafunc}
\ee
Let $w_*=w_*(d_{\rm SD};\bar{\mu}_a,D_0,\mathfrak{n})$ be the positive number such that $\Lambda(w_*;a,b)=0$. Then the depth of the banana is obtained as
\begin{equation}
z_0=\frac{d_{\rm SD}}{2}w_*.
\label{z0value}
\end{equation}

In the case that $\bar{\mu}_a$ is negligibly small, we can set $a=0$. Then for large $w$,
\be
\Lambda(w;0,b)=
\int_0^{\infty}J_0(x)\frac{x^2e^{-wx}}{1+bx}\,dx\approx
\int_0^{\infty}x^2e^{-wx}\,dx=
\frac{2}{w^3}>0.
\ee
When $w=0$, we have
\ba
\Lambda(0;0,b)
&=
\int_0^{\infty}J_0(x)\frac{x^2}{1+bx}\,dx
\\
&=
\frac{1}{b}\int_0^{\infty}J_0(x)x\,dx-\frac{1}{b}\int_0^{\infty}J_0(x)\frac{x}{1+bx}\,dx
\\
&=
-\frac{1}{b^3}\int_0^{\infty}J_0(t/b)\frac{t}{1+t}\,dt
\\
&=
-\frac{1}{b^2}+\frac{\pi}{2b^3}\left(H_0\left(\frac{1}{b}\right)-Y_0\left(\frac{1}{b}\right)\right)<0,
\ea
where we used the Hankel transform $\int_0^{\infty}J_0(qx)x\,dx=\delta(q)/q$. Here, the Struve function $H_0$ and Bessel function of the second kind $Y_0$ are given by
\ba
H_0(x)
&=
\frac{1}{\sqrt{\pi}}\sum_{n=0}^{\infty}\frac{(-1)^nx^{2n+1}}{[(2n+1)!!]2^n\Gamma(n+3/2)},
\\
Y_0(x)
&=
\frac{1}{\pi}\left.\left(\frac{\pp J_{\nu}(x)}{\pp\nu}-\frac{\pp J_{-\nu}(x)}{\pp\nu}\right)\right|_{\nu=0}
\\
&=
\frac{2}{\pi}J_0(x)\ln\frac{x}{2}-\frac{2}{\pi}\sum_{n=0}^{\infty}\frac{(-1)^n}{(n!)^2}\left(\frac{x}{2}\right)^{2n}\psi(n+1),
\ea
where $\psi$ is the digamma function. The facts that $\Lambda(w;0,b)$ is positive for large $w$ and negative for small $w>0$ imply that $\Lambda(w;0,b)$ has a zero.

If in addition, $b=0$ (i.e., the zero boundary condition $z_e=0$), we obtain
\begin{equation}
\Lambda(w;0,0)=\int_0^{\infty}J_0(x)x^2e^{-wx}\,dx=
\frac{\sqrt{w^2+1}(2w^2-1)}{(1+w^2)^3}.
\label{lamfunc0}
\end{equation}
Hence $w_*=1/\sqrt{2}$. In the case that $\bar{\mu}_a$ is negligiblly small and $z_e=0$, the depth is given by $z_0=d_{\rm SD}/(2\sqrt{2})\approx0.35d_{\rm SD}.$

\begin{figure}[htbp]
\centering
\includegraphics[width=0.45\textwidth]{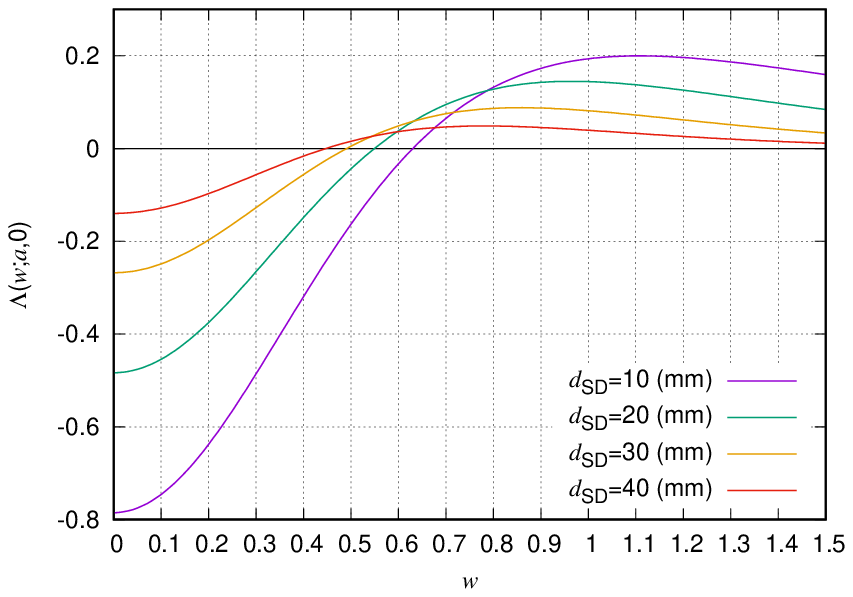}
\includegraphics[width=0.45\textwidth]{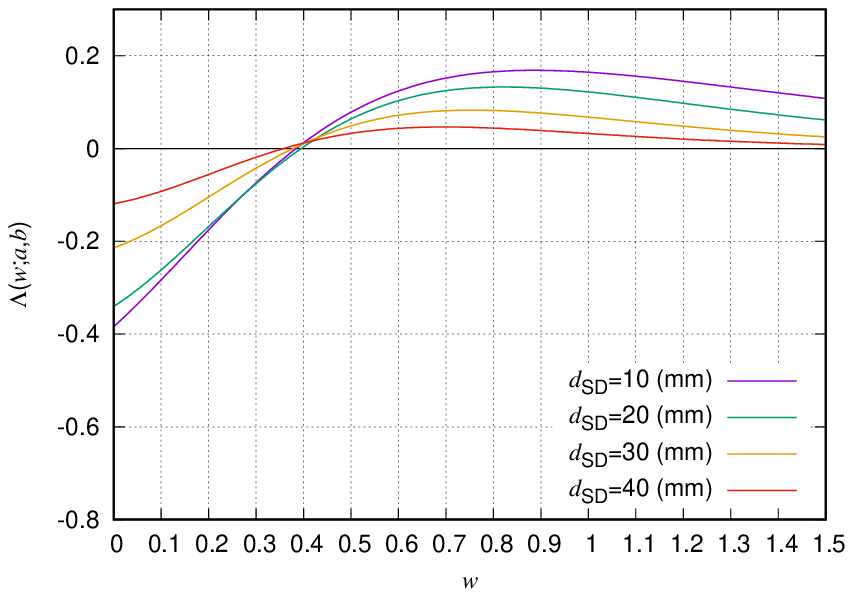}
\caption{
(Left) The function $\Lambda(w;a,0)$ is plotted for different $d_{\rm SD}$. (Right) The function $\Lambda(w;a,b)$ is plotted for different $d_{\rm SD}$ for $\mathfrak{n}=1.4$.
}
\label{fig_lambzero}
\end{figure}

\begin{figure}[htbp]
\centering
\includegraphics[width=0.45\textwidth]{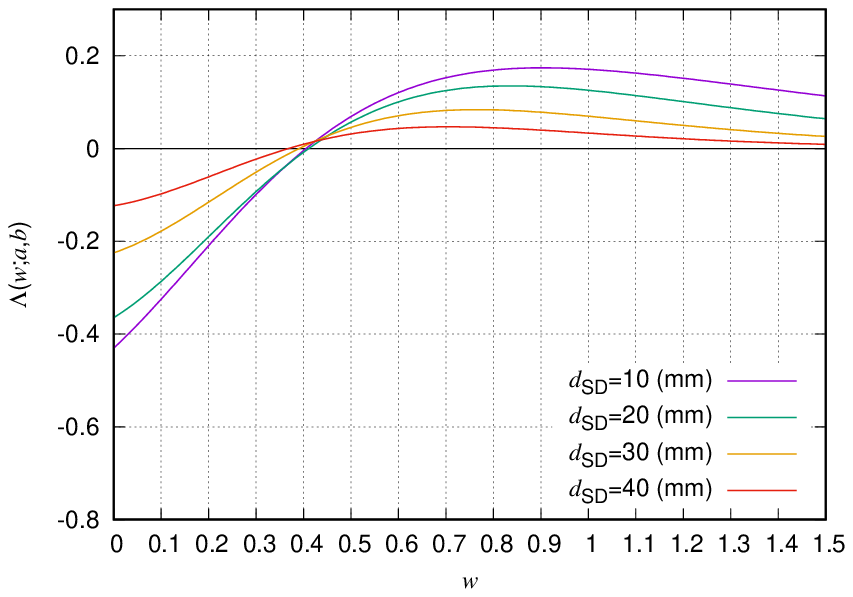}
\includegraphics[width=0.45\textwidth]{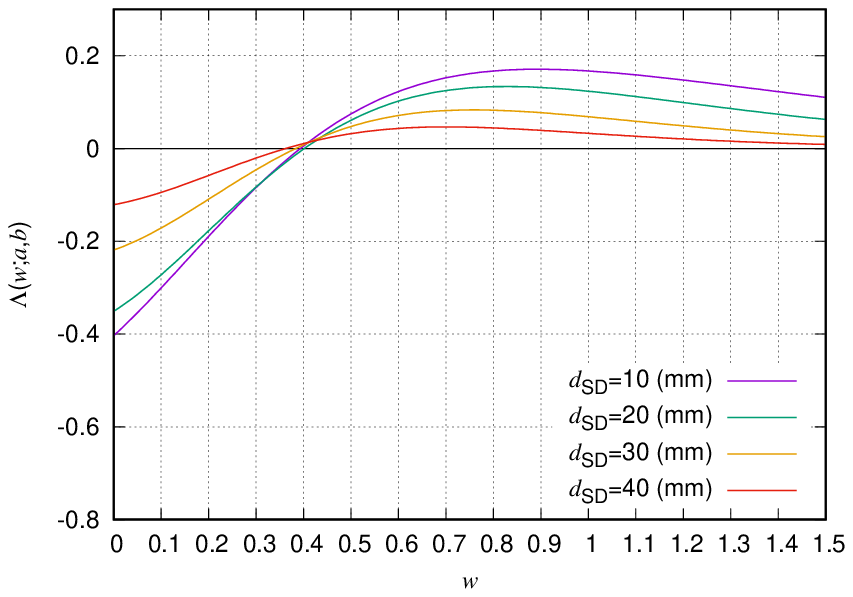}
\caption{
The function $\Lambda(w;a,b)$ is plotted for different $d_{\rm SD}$ for (Left) $\mathfrak{n}=1.33$ and (Right) $\mathfrak{n}=1.37$.
}
\label{fig_lamb133}
\end{figure}

Let us find $w_*$ for general cases by plotting (\ref{Lambdafunc}). To numerically handle the oscillatory integral in (\ref{Lambdafunc}), we changed the variable as $y=\sqrt{x^2-a^2}$. Let us assume that $\bar{\mu}_a=0.01\,{\rm mm}^{-1}$ and $\mu_s'=1\,{\rm mm}^{-1}$. That is, $a\approx0.0866 d_{\rm SD}$. In Figs.~\ref{fig_lambzero} and \ref{fig_lamb133}, $\Lambda(w;a,b)$ is plotted against $w$ for different $d_{\rm SD}$. The zero boundary condition ($z_e=0$) was assumed in the left panel of Fig.~\ref{fig_lambzero}. In Fig.~\ref{fig_lambzero} (Left), $w_*=0.49$ for $d_{\rm SD}=30\,{\rm mm}$. In the case of the zero boundary condition, the depth of the banana is estimated as $z_0=7.5\,{\rm mm}$. The right panel of Fig.~\ref{fig_lambzero} shows $\Lambda(w;a,b)$ in the case of $\mathfrak{n}=1.4$. We found $w_*=0.38$ for $d_{\rm SD}=30\,{\rm mm}$ and $z_0=5.7\,{\rm mm}$ in Fig.~\ref{fig_lambzero} (Right). In Fig.~\ref{fig_lamb133}, (Left) $\mathfrak{n}=1.33$ and (Right) $\mathfrak{n}=1.37$. In both panels in Fig.~\ref{fig_lamb133}, $w_*=0.39$ for $d_{\rm SD}=30\,{\rm mm}$ and $z_0=5.9\,{\rm mm}$. We found $w_*\approx0.4$ for the Robin boundary condition in Fig.~\ref{fig_lambzero} (Right), and Fig.~\ref{fig_lamb133} (Left) and (Right). That is, we have an approximate relation:
\be
z_0\approx0.2d_{\rm SD}.
\ee
Zeros $w_*$ found in Figs.~\ref{fig_lambzero} and \ref{fig_lamb133} are summarized in Tables \ref{table1}.

\begin{table}[htbp]
\centering
\begin{tabular}{|c|c|}
\hline
$z_e=0$ & \\
\hline
$d_{\rm SD}$ [${\rm mm}$] & $w_*$ \\
\hline
10 & 0.63 \\
20 & 0.55 \\
30 & 0.49 \\
40 & 0.45 \\
\hline
\end{tabular}
\begin{tabular}{|c|c|}
\hline
$\mathfrak{n}=1.4$ & \\
\hline
$d_{\rm SD}$ [${\rm mm}$] & $w_*$ \\
\hline
10 & 0.38 \\
20 & 0.40 \\
30 & 0.38 \\
40 & 0.36 \\
\hline
\end{tabular}
\begin{tabular}{|c|c|}
\hline
$\mathfrak{n}=1.33$ & \\
\hline
$d_{\rm SD}$ [${\rm mm}$] & $w_*$ \\
\hline
10 & 0.41 \\
20 & 0.41 \\
30 & 0.39 \\
40 & 0.37 \\
\hline
\end{tabular}
\begin{tabular}{|c|c|}
\hline
$\mathfrak{n}=1.37$ & \\
\hline
$d_{\rm SD}$ [${\rm mm}$] & $w_*$ \\
\hline
10 & 0.40 \\
20 & 0.40 \\
30 & 0.39 \\
40 & 0.36 \\
\hline
\end{tabular}
\caption{
The relation between $d_{\rm SD}$ and $w_*$ is shown in four tables, from the left, for $z_e=0$ (zero boundary condition) read from Fig.~\ref{fig_lambzero} (Left), $\mathfrak{n}=1.4$ read from Fig.~\ref{fig_lambzero} (Right), $\mathfrak{n}=1.33$ read from Fig.~\ref{fig_lamb133} (Left), and $\mathfrak{n}=1.37$ read from Fig.~\ref{fig_lamb133} (Right).
}
\label{table1}
\end{table}

\begin{figure}[htbp]
\centering
\includegraphics[width=0.45\textwidth]{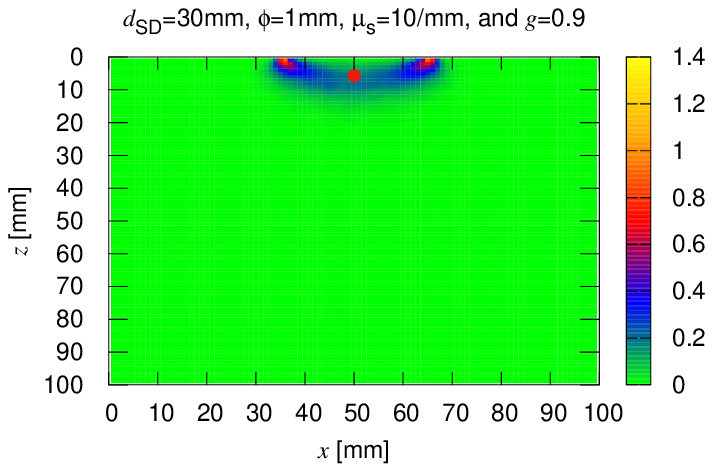}
\includegraphics[width=0.45\textwidth]{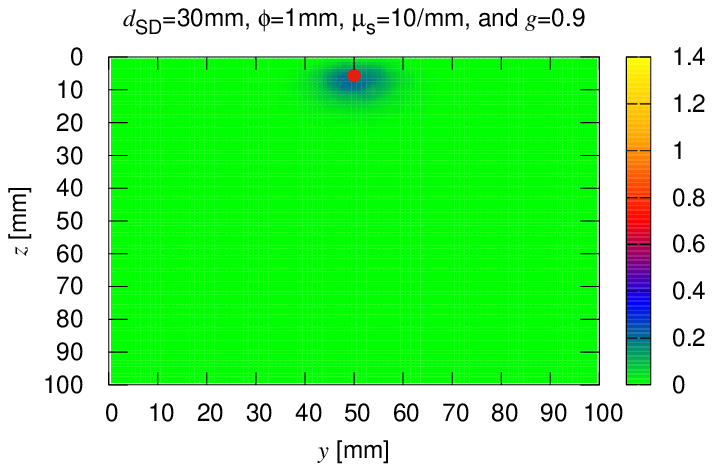}
\caption{
The spatial sensitivity profile is shown for $d_{\rm SD}=30\,{\rm mm}$, $\phi=1\,{\rm mm}$, $\mu_s=10\,{\rm mm}^{-1}$, and ${\rm g}=0.9$. The cross sections at (Left) $y=50\,{\rm mm}$ and (Right) $x=50\,{\rm mm}$ are shown. The point at $z=z_0$ is shown by a solid red circle.
}
\label{mcfig1}
\end{figure}

\begin{figure}[htbp]
\centering
\includegraphics[width=0.45\textwidth]{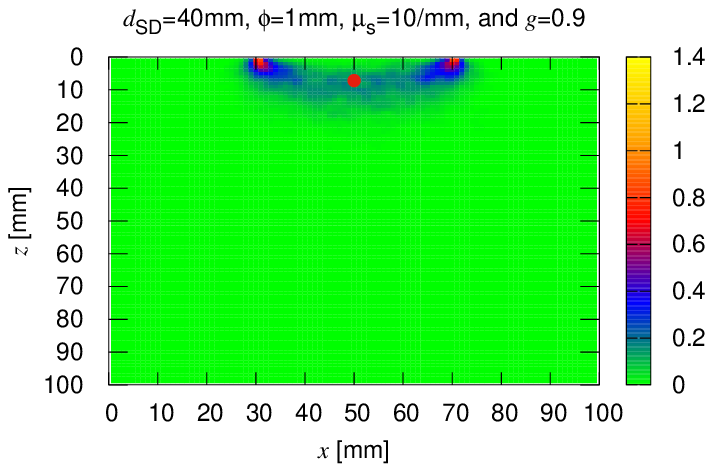}
\includegraphics[width=0.45\textwidth]{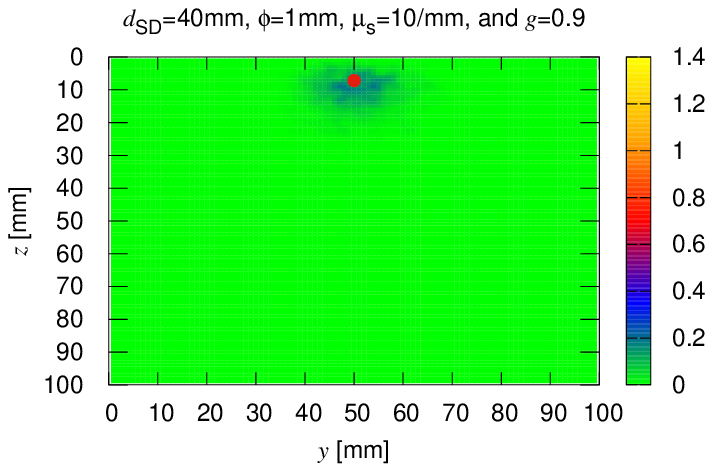}
\caption{
The spatial sensitivity profile is shown for $d_{\rm SD}=40\,{\rm mm}$, $\phi=1\,{\rm mm}$, $\mu_s=10\,{\rm mm}^{-1}$, and ${\rm g}=0.0$. The cross sections at (Left) $y=50\,{\rm mm}$ and (Right) $x=50\,{\rm mm}$ are shown.
}
\label{mcfig2}
\end{figure}

In Figs.~\ref{mcfig1} and \ref{mcfig2}, we plot $z_0$ with the spatial sensitivity profile \cite{Arridge95a,Arridge-Schweiger95,Okada-Delpy03a}, which was obtained by Monte Carlo simulation. Let $\phi,{\rm g}$ be the diameter of an optical fiber and scattering asymmetry parameter, respectively. We set $\bar{\mu}_a=0.01\,{\rm mm}^{-1}$. We note $\mu_s'=(1-{\rm g})\mu_s=1\,{\rm mm}^{-1}$, where $\mu_s$ is the scattering coefficient. The source and detector were placed on the $x$-axis. In both Figs.~\ref{mcfig1} and \ref{mcfig2}, planes at (Left) $y=50\,{\rm mm}$ and (Right) $x=50\,{\rm mm}$ are shown. The center position of the banana at $z_0$ is shown by a solid red circle in both cases of $d_{\rm SD}=30\,{\rm mm}$ and $d_{\rm SD}=40\,{\rm mm}$. In Monte Carlo simulation, the refractive index of the medium was set to $\mathfrak{n}=1.4$ and the Fresnel reflection was assumed on the boundary.

More results from Monte Carlo simulation are shown in Appendix \ref{mcsimulation}. Spatial sensitivity profiles are shown for 
(Fig.~\ref{mcfig3010}) $d_{\rm SD}=30\,{\rm mm}$, $\phi=1\,{\rm mm}$, $\mu_s=1\,{\rm mm}^{-1}$, ${\rm g}=0.0$, 
(Fig.~\ref{mcfig3059}) $d_{\rm SD}=30\,{\rm mm}$, $\phi=5\,{\rm mm}$, $\mu_s=10\,{\rm mm}^{-1}$, ${\rm g}=0.9$,
(Fig.~\ref{mcfig3050}) $d_{\rm SD}=30\,{\rm mm}$, $\phi=5\,{\rm mm}$, $\mu_s=1\,{\rm mm}^{-1}$, ${\rm g}=0.0$,
(Fig.~\ref{mcfig4010}) $d_{\rm SD}=40\,{\rm mm}$, $\phi=1\,{\rm mm}$, $\mu_s=1\,{\rm mm}^{-1}$, ${\rm g}=0.0$,
(Fig.~\ref{mcfig4059}) $d_{\rm SD}=40\,{\rm mm}$, $\phi=5\,{\rm mm}$, $\mu_s=10\,{\rm mm}^{-1}$, ${\rm g}=0.9$,
(Fig.~\ref{mcfig4050}) $d_{\rm SD}=40\,{\rm mm}$, $\phi=5\,{\rm mm}$, $\mu_s=1\,{\rm mm}^{-1}$, ${\rm g}=0.0$.

We note that the depth $z_0$ of the banana in (\ref{z0value}) is different from the mean photon-visit depth $\langle z\rangle_{d_{\rm SD}}$ and the reciprocal $\delta$ of the decay rate $\mu_{\rm eff}$, which are given by\cite{Patterson-etal95}
\begin{equation}
\langle z\rangle_{d_{\rm SD}}=\frac{1}{2}\sqrt{d_{\rm SD}\delta},\quad
\delta=\frac{1}{\mu_{\rm eff}},
\label{zsdformula}
\end{equation}
where
\begin{equation}
\mu_{\rm eff}=\sqrt{3\mu_a(\mu_a+\mu_s')}.
\label{mueff}
\end{equation}
In the case of $\mu_a=0.01\,{\rm mm}^{-1}$, $\mu_s'=1\,{\rm mm}^{-1}$, and $\mathfrak{n}=1.4$, we have $z_0=5.7\,{\rm mm}$, $\langle z\rangle_{d_{\rm SD}}=6.56\,{\rm mm}$, $\delta=5.74\,{\rm mm}$ for $d_{\rm SD}=30\,{\rm mm}$, and $z_0=7.2\,{\rm mm}$, $\langle z\rangle_{d_{\rm SD}}=7.58\,{\rm mm}$, $\delta=5.74\,{\rm mm}$ for $d_{\rm SD}=40\,{\rm mm}$. We note that the formula (\ref{zsdformula}) was phenomenologically derived assuming $\mathfrak{n}\approx1.36$.

The depth $z_0$ of the banana can be compared with the exponential decay of the spatially oscillating illumination with $q_0$ (i.e., frequency is $f=q_0/(2\pi)$). The specific intensity decays as $I\sim e^{-z/\xi_{q_0}}$, where $\xi_{q_0}=1/\sqrt{3\mu_a\mu_s'+q_0^2}$ for small $\mu_a>0$ \cite{Cuccia-etal09,Machida-etal20}. For example, when $\mu_s'=1\,{\rm mm}^{-1}$, $\mu_a=0.01\,{\rm mm}^{-1}$, $q_0=1\,{\rm mm}^{-1}$ (i.e., $f=0.16\,{\rm mm}^{-1}$), we have $\xi_{q_0}\approx 1\,{\rm mm}$. Since $z_0=5.7\,{\rm mm}$ for $d_{\rm SD}=30\,{\rm mm}$, we see $z_0\gg\xi_{q_0}$. This motivates us to consider the stripe illumination, which is a structured illumination with the combination of spatial-frequency domain imaging and point-illumination imaging.

\subsection{Impulse stripe illumination}
\label{psi}

\begin{figure}[htbp]
\centering
\includegraphics[width=0.8\textwidth]{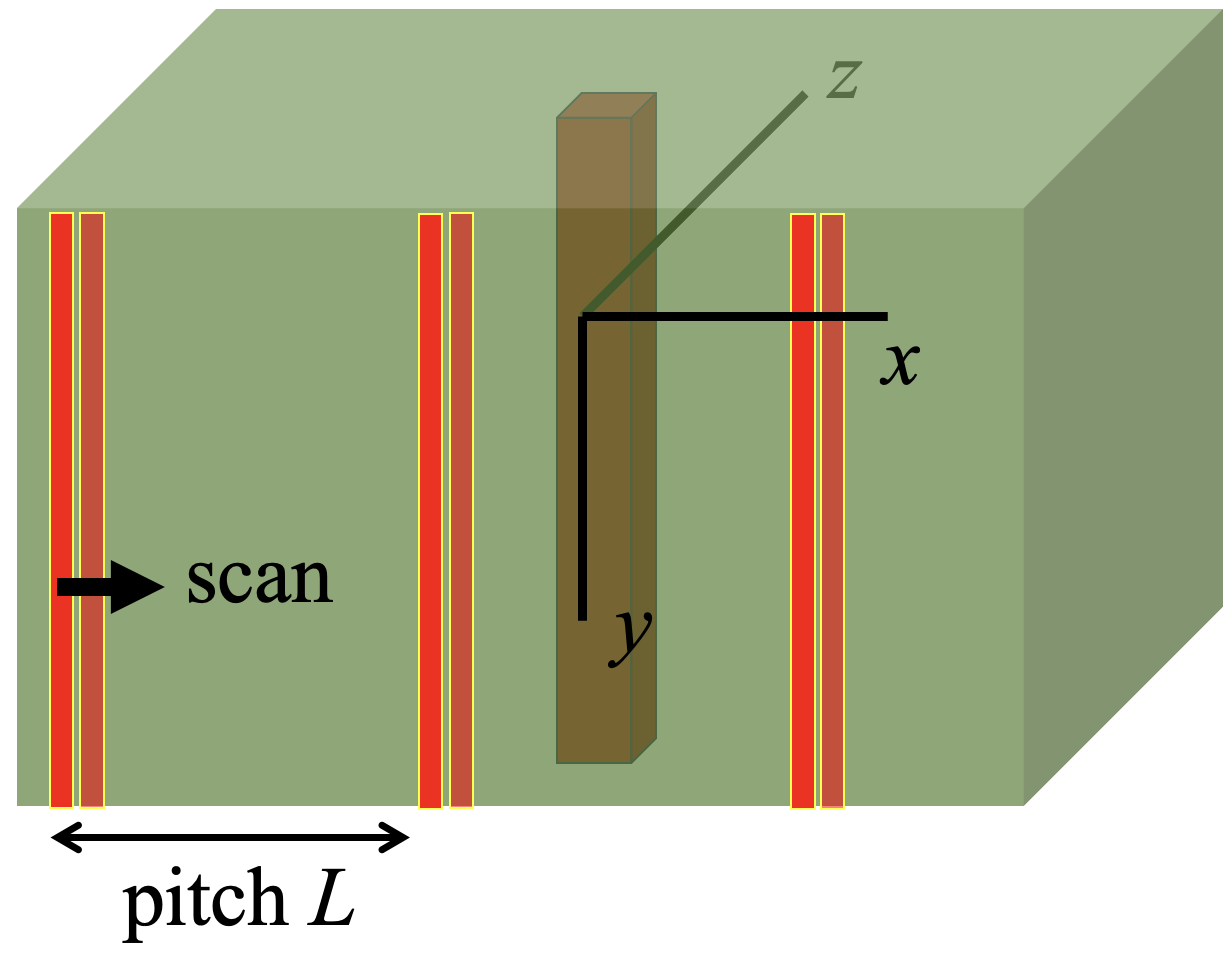}
\caption{
Schematic figure for the stripe illumination of pitch $L$. The same illumination pattern is scanned in the positive $x$ direction with step $\ell$. The absorption inhomogeneity $\delta\mu_a(\rvv{r})$ shown in the domain is the absorption bar assumed in Sec.~\ref{DOT}. In the $x$-$z$ plane, the bar has four corners at $(-6,4)$, $(-6,10)$, $(0,4)$, and $(0,10)$ in the unit of ${\rm mm}$.
}
\label{fig_dot}
\end{figure}

Let $c_0$ be the speed of light in vacuum. Then $c=c_0/\mathfrak{n}$ is the speed of light in the medium. Let $T>0$ be the observation time. The diffuse fluence rate $u(\rvv{r},t)$ for the impulse illumination obeys the following diffusion equation.
\be
\left\{\begin{aligned}
\frac{1}{c}\frac{\pp}{\pp t}u-D_0\Delta u+\mu_a(\rvv{r})u=f,
&\quad(x,t)\in\Omega\times(0,T),
\\
-D_0\frac{\pp}{\pp z}u+\frac{1}{\zeta}u=0,
&\quad(x,t)\in\pp\Omega\times(0,T),
\\
u=0,
&\quad\rvv{r}\in\Omega,\quad t=0,
\end{aligned}\right.
\label{de1}
\ee
where $f(\rvv{r},t)$ is the incident beam. The absorption coefficient $\mu_a(\rvv{r})$ is not necessarily a point absorber and we assume that $\mu_a(\rvv{r})$ is given by
\be
\mu_a(\rvv{r})=\bar{\mu}_a+\delta\mu_a(\rvv{r}) \ ,
\ee
where $\bar{\mu}_a$ is a positive constant and $\left.\mu_a\right|_{\pp\Omega}=\bar{\mu}_a$. The source term $f(\rvv{r},t)$ is given by
\be
f(\rvv{r},t)=f_0s(\rvv{\rho})\delta(t)\delta(z),
\ee
where $f_0>0$ is a constant which is determined by the beam. Let $L,\ell$ be the pitch and scan step, respectively. A schematic figure of the illumination is shown in Fig.~\ref{fig_dot}. The function $s(\rvv{\rho})$ is given by
\begin{equation}
s(\rvv{\rho})=s_n(\rvv{\rho})=
\sum_{j=-\infty}^{\infty}\delta\left(x-jL-\frac{2n-1}{2}\ell\right),
\label{afunc}
\end{equation}
where $n=1,\dots,N_f$ and $N_f$ is the number of scans. For illumination, we set
\be
L=32\,{\rm mm},\quad\ell=2\,{\rm mm},\quad N_f=16.
\ee

We define
\be
p_l=\frac{2\pi l}{L},\quad l\in\Zm.
\ee
Then the function $s_n(\rvv{\rho})$ can be written as
\begin{equation}
s_n(\rvv{\rho})=
\frac{1}{L}\sum_{l=-\infty}^{\infty}e^{ip_lx}e^{-ip_l(2n-1)\ell/2},
\label{wdecom}
\end{equation}
where we used the Poisson sum formula,
\begin{equation}
\sum_{j=-\infty}^{\infty}\delta(x-2\pi j)=
\frac{1}{2\pi}\sum_{l=-\infty}^{\infty}e^{ilx}.
\label{poissonsum}
\end{equation}

The expression (\ref{wdecom}) implies that the solution to (\ref{de1}) can be expressed as
\begin{equation}
u(\rvv{r},t)=\sum_{l=-\infty}^{\infty}u_l(\rvv{r},t),
\label{udecom}
\end{equation}
where $u_l$ ($l=-\infty,\dots,\infty$) is the solution of (\ref{de1}) for
\be
f(\rvv{r},t)=f_l(\rvv{r},t)=
\frac{f_0}{L}e^{ip_lx}e^{-ip_l(2n-1)\ell/2}\delta(t)\delta(z).
\ee

As discussed in the end of Sec.~\ref{banana}, for constant $\mu_a$, it is known that the penetration depth (i.e., the decay rate of $u(\rvv{r})$ in the $z$-direction) depends on the spatial frequency $p_l$. We have asymptotically \cite{Cuccia-etal09,Machida-etal20}
\be
u_l(\rvv{r},t)\sim\mbox{const.}\times e^{-z\sqrt{\mu_{\rm eff}^2+p_l^2}},
\ee
where $\mu_{\rm eff}$ is given in (\ref{mueff}). Since $u$ is given by the sum of $u_l$ in (\ref{udecom}), the diffuse light $u$ penetrates deeper than the light by a single sinusoidal illumination.

Next, we consider how deep light from the stripe illumination of the pitch $L$ can penetrate the medium. Let us look at the outgoing light which is detected at $\rvv{r}_d^{(m,n)}={^t}(x_d^{(m,n)},y,0)$, where $y\in\Rm$ and
\begin{equation}
x_d^{(m,n)}=\frac{2m-1}{2}L+(n-1)\ell,\quad m\in\Zm
\label{m2:detectpoints}
\end{equation}
for each $n=1,\dots,N_f$. To discuss the typical penetration depth, let us consider the separation between the source at $(2n-1)\ell/2$ and $x_d^{(1,n)}$ since the outgoing light at multiple pitches away is negligible. This distance $d_{\rm SD}$ is given by
\be
d_{\rm SD}=\frac{L-\ell}{2}=15\,{\rm mm},
\ee
where we put $L=32\,{\rm mm}$ and $\ell=2\,{\rm mm}$. Assuming the relation $z_0\approx 0.2d_{\rm SD}$, we find that the depth of the banana is
\begin{equation}
z_0\approx 3\,{\rm mm}.
\label{z03mm}
\end{equation}
The above formula $d_{\rm SD}=(L-\ell)/2$ implies that the information at a deep tissue can be extracted with large $L$.

\subsection{Diffuse optical tomography}
\label{DOT}

The forward data were taken numerically for diffuse optical tomography. For this purpose, the Monte Carlo eXtreme (MCX) was used \cite{Fang-Boas09}. The number of photons used in each simulation was $10^8$. In our numerical experiment by Monte Carlo simulation, an absorption bar ($\mu_a=0.02\,{\rm mm}^{-1}$, $\mu_s=10\,{\rm mm}^{-1}$, ${\rm g}=0.9$) is embedded along the $y$-axis. The cross section of the bar is parallel to the $x$-$z$ plane and is a square of side length $6\,{\rm mm}$. The depth of the center of the square is $7\,{\rm mm}$. In the $x$-$z$ plane, the positions of four corners of the absorber rectangle are $(-6,4)$, $(-6,10)$, $(0,4)$, and $(0,10)$ in the unit of ${\rm mm}$. In the depth direction along the $z$-axis, the absorption bar exists in $4\,{\rm mm}\le z\le 10\,{\rm mm}$. See Fig.~\ref{fig_dot}. The Monte Carlo simulation was performed in a box ($-64\le x\le 64\,{\rm mm}$, $-64\le y\le 64\,{\rm mm}$, $0\le z\le 40\,{\rm mm}$) with voxel size $2\,{\rm mm}\,\times\,2\,{\rm mm}\,\times\,2\,{\rm mm}$. The Robin boundary condition is considered on the illumination plane (i.e., according to the Fresnel reflection a part of outgoing photons reenters the medium instead of exiting it). The zero boundary condition is imposed on other boundaries. In the medium, we set
\be
\bar{\mu}_a=0.01\,{\rm mm}^{-1},\quad
\mu_s'=1\,{\rm mm}^{-1},\quad\mathfrak{n}=1.4.
\ee
We remark that the target shape in the $y$-direction is assumed to be known in this optical tomography. That is, $\delta\mu_a$ is independent of $y$ and the reconstruction in the $x$-$z$ plane will be performed. Furthermore, the depth of the target is slightly deeper than the center $z_0$ of the banana given in (\ref{z03mm}), so that the deeper half of the banana covers the bar target. The target depth is deeper than the depth typically assumed in spatial frequency domain imaging \cite{Cuccia-etal05}.

Let us extend the interval $(0,T)$ to $(-\infty,\infty)$ by the limit $T\to\infty$ and zero extension for $t<0$, we consider the Fourier transform as
\be
v^n(\rvv{r},\omega)=\int_{-\infty}^{\infty}e^{-i\omega t}u(\rvv{r},t)\,dt
\ee
for $n=1,\dots,N_f$.

We have
\be
\left\{\begin{aligned}
-D_0\Delta v^n+\left(\alpha(\omega)+\delta\mu_a(\rvv{r})\right)v^n=f_0a_n(\rvv{\rho})\delta(z),
&\quad x\in\Omega,
\\
-D_0\frac{\pp}{\pp z}v^n+\frac{1}{\zeta}v^n=0,
&\quad x\in\pp\Omega,
\end{aligned}\right.
\label{de2}
\ee
where
\be
\alpha(\omega)=\bar{\mu}_a+\frac{i\omega}{c}.
\ee

Let us consider the Green's function $G_{\omega}(\rvv{r},\rvv{r}')$ for (\ref{de2}), which satisfies
\be
\left\{\begin{aligned}
-D_0\Delta G_{\omega}+\alpha(\omega)G_{\omega}=\delta(\rvv{r}-\rvv{r}'),
&\quad x\in\Omega,
\\
-D_0\frac{\pp}{\pp z}G_{\omega}+\frac{1}{\zeta}G_{\omega}=0,
&\quad x\in\pp\Omega.
\end{aligned}\right.
\label{green2}
\ee
We obtain
\begin{equation}
\begin{aligned}
G_{\omega}(\rvv{r},\rvv{r}')
&=
\frac{1}{(2\pi)^2}\int_{\Rm^2}e^{i\rvv{q}\cdot(\rvv{\rho}-\rvv{\rho}')}
\widetilde{G}_{\omega}(|\rvv{q}|,z,z')\,d\rvv{q},
\end{aligned}
\label{gwdef}
\end{equation}
where
\be
Q_{\omega}(q)=\sqrt{\frac{\alpha(\omega)}{D_0}+q^2},
\ee
and
\begin{equation}
\widetilde{G}_{\omega}(q,z,z')=
\frac{1}{2D_0Q_{\omega}(q)}
\left[e^{-Q_{\omega}(q)|z-z'|}-\frac{1-Q_{\omega}(q)z_e}{1+Q_{\omega}(q)z_e}e^{-Q_{\omega}(q)|z+z'|}\right]
\label{Homedef}
\end{equation}
for $q\in\Rm$.

Let $v_0^n$ be the solution of the diffusion equation in (\ref{de2}) for which $\delta\mu_a$ is removed. We have
\ba
v_0^n(\rvv{r},\omega)
&=
f_0\sum_{j=-\infty}^{\infty}\int_{-\infty}^{\infty}G_{\omega}\left(\rvv{r};jL+\frac{2n-1}{2}\ell,y',0\right)\,dy'
\\
&=
\frac{f_0}{L}\sum_{l=-\infty}^{\infty}\widetilde{G}_{\omega}(p_l,z,0)
e^{ip_lx}e^{-ip_l(2n-1)\ell/2}
\\
&=
v_0^n(x,z,\omega),
\ea
where the Green's function was expressed as $G_{\omega}(\rvv{r},\rvv{r}')=G_{\omega}(\rvv{r};x',y',z')$.

We give detection points as
\be
x_i=(i-1)h_d+x_{\rm init}\quad(i=1,\dots,N_d).
\ee
We will set
\be
N_d=33,\quad h_d=2\,{\rm mm},\quad x_{\rm init}=-32\,{\rm mm}.
\ee
That is, $-32\,{\rm mm}\le x_i\le 32\,{\rm mm}$. We write $\rvv{r}_d^{(i)}={^t}(x_i,0,0)$. Within the diffusion approximation we have
\be
\ln{\frac{v_0^n(\rvv{r}_d^{(i)},\omega)}{v^n(\rvv{r}_d^{(i)},\omega)}}
\approx
\ln{\frac{I_0^n(\rvv{r}_d^{(i)},-\hvv{z},\omega)}{I^n(\rvv{r}_d^{(i)},-\hvv{z},\omega)}},
\ee
where $\hvv{z}={^t}(0,0,1)$. Here, $I_0^n(\rvv{r}_d^{(i)},-\hvv{z},\omega)$ and $I^n(\rvv{r}_d^{(i)},-\hvv{z},\omega)$ are Fourier transforms of specific intensities in the outer normal direction at $\rvv{r}_d^{(i)}$ for $\mu_a=\bar{\mu}_a$ and $\mu_a=\bar{\mu}_a+\delta\mu_a$ with the $n$th scan.

We note that
\begin{equation}
\begin{aligned}
&
\int_{-\infty}^{\infty}G_{\omega}(\rvv{r},\rvv{r}')\,dy'
\\
&=
\frac{1}{2\pi}\int_{-\infty}^{\infty}e^{iq(x-x')}
\frac{1}{2D_0Q_{\omega}(q)}
\left[e^{-Q_{\omega}(q)|z-z'|}-\frac{1-Q_{\omega}(q)z_e}{1+Q_{\omega}(q)z_e}e^{-Q_{\omega}(q)|z+z'|}\right]\,dq
\\
&=
\frac{1}{\pi}\int_0^{\infty}\cos[q(x-x')]\frac{1}{2D_0Q_{\omega}(q)}
\left[e^{-Q_{\omega}(q)|z-z'|}-\frac{1-Q_{\omega}(q)z_e}{1+Q_{\omega}(q)z_e}e^{-Q_{\omega}(q)|z+z'|}\right]\,dq
\\
&=
H_{\omega}(x-x',z,z').
\end{aligned}
\label{m2:Homedef}
\end{equation}
The numerical computation of $H_{\omega}$ is described in Appendix \ref{m2:doubleexp}.

For each $n$, light is detected at $N_d$ points. We define
\be
\psi^n(x_i,\omega)=
v_0^n(x_i,0,0,\omega)
\ln{\frac{\left\langle I_0^n(\rvv{r}_d^{(i)},-\hvv{z},\omega)\right\rangle}{\left\langle I^n(\rvv{r}_d^{(i)},-\hvv{z},\omega)\right\rangle}},
\quad i=1,\dots,N_d,
\ee
where $\langle\cdot\rangle$ denotes the average over $y$.

Since $\delta\mu_a$ does not depend on $y$, we write
\be
\delta\mu_a(\rvv{r})=\eta(x,z).
\ee
We note the identity:
\be
v^n(\rvv{r},\omega)=
v_0^n(\rvv{r},\omega)-
\int_{\Omega}G_{\omega}(\rvv{r},\rvv{r}')\delta\mu_a(\rvv{r}')v^n(\rvv{r}',\omega)\,d\rvv{r}'.
\ee
Since $v_0^n$ is independent of $y$, the above identity implies $v^n(\rvv{r},\omega)=v^n(x,z,\omega)$ is also independent of $y$. We have
\be
v^n(x,z,\omega)=
v_0^n(x,z,\omega)
-\int_0^{\infty}\int_{-\infty}^{\infty}
H_{\omega}(x-x',z,z')\eta(x',z')v^n(x',z',\omega)\,dx'dz'.
\ee
By the (first) Born approximation,
\be
v^n\approx v_0^n+v_1^n,
\ee
where
\be
v_1^n(x,z,\omega)
=-\int_0^{\infty}\int_{-\infty}^{\infty}v_0^n(x',z',\omega)
H_{\omega}(x-x',z,z')\eta(x',z')\,dx'dz'.
\ee

Let us write
\be
\left(K\eta\right)(x,\omega)=-\sum_{n=1}^{N_f}v_1^n(x,0,\omega).
\ee
We have in the (first) Rytov approximation:
\be
\sum_{n=1}^{N_f}\psi^n(x_i,\omega)=(K\eta)(x_i,\omega),\quad i=1,\dots,N_d.
\ee
By the Fourier transform and summing over $n$, we introduce
\ba
\Psi(q,\omega)
&=
\sum_{n=1}^{N_f}\int_{-\infty}^{\infty}e^{-iqx}\psi^n(x,\omega)\,dx
\\
&\approx
\sum_{n=1}^{N_f}\int_{x_1}^{x_{N_d}}e^{-iqx}\psi^n(x,\omega)\,dx
.
\ea
Similarly we can introduce
\be
(\widetilde{K}\eta)(q,\omega)=\int_{-\infty}^{\infty}e^{-iqx}(K\eta)(x,\omega)\,dx.
\ee
We have
\ba
(\widetilde{K}\eta)(q,\omega)
&=
\sum_{n=1}^{N_f}\int_{-\infty}^{\infty}e^{-iqx}
\int_0^{\infty}\int_{-\infty}^{\infty}v_0^n(x',z',\omega)
H_{\omega}(x-x',0,z')\eta(x',z')\,dx'dz'dx
\\
&\approx
\frac{1}{2}\sum_{n=1}^{N_f}
\int_0^{\infty}\int_{-\infty}^{\infty}
e^{-iqx'}\frac{v_0^n(x',z',\omega)}{D_0Q_{\omega}(q_d)}
\\
&\times
\left[e^{-Q_{\omega}(q)z'}-\frac{1-Q_{\omega}(q)z_e}{1+Q_{\omega}(q)z_e}e^{-Q_{\omega}(q)z'}\right]
\eta(x',z')\,dx'dz'
\\
&=
\frac{f_0}{L}\sum_{n=1}^{N_f}\sum_{l=-\infty}^{\infty}\int_0^{\infty}
\widetilde{G}_{\omega}(p_l,z',0)e^{-ip_l(2n-1)\ell/2}
\\
&\times
\widetilde{G}_{\omega}(q,0,z')\widetilde{\eta}(q-p_l,z')\,dz',
\ea
where we defined
\be
\widetilde{\eta}(q,z)=\int_{-\infty}^{\infty}e^{-iqx}\eta(x,z)\,dx.
\ee
Recalling $L=N_f\ell$, we can further proceed as
\begin{equation}\begin{aligned}
(\widetilde{K}\eta)(q,\omega)
&=
\frac{f_0N_f}{L}\sum_{l=mN_f,\;m\in\Zm}\int_0^{\infty}e^{ip_l\ell/2}
\widetilde{G}_{\omega}(p_l,z',0)\widetilde{G}_{\omega}(q,0,z')
\widetilde{\eta}(q-p_l,z')\,dz'
\\
&=
\frac{f_0N_f}{L}\sum_{m=-\infty}^{\infty}\int_0^{\infty}e^{i\pi m}
\widetilde{G}_{\omega}\left(\frac{2\pi m}{\ell},z',0\right)
\widetilde{G}_{\omega}(q,0,z')
\widetilde{\eta}\left(q-\frac{2\pi m}{\ell},z'\right)\,dz'
\\
&\approx
\frac{f_0N_f}{L}\int_0^{\infty}
\widetilde{G}_{\omega}(0,z',0)\widetilde{G}_{\omega}(q,0,z')
\widetilde{\eta}\left(q,z'\right)\,dz'.
\end{aligned}
\label{tildeK}
\end{equation}
Thus we arrive at the following linear problem.
\be
\Psi(q,\omega)=
\int_0^{\infty}\left[\frac{f_0N_f}{h_dL}
\widetilde{G}_{\omega}(0,z',0)\widetilde{G}_{\omega}(q,0,z')\right]
\widetilde{\eta}(q,z')\,dz'.
\ee

We can discretize $\omega,z$ as
\be
\omega_k=(k-1)\Delta\omega\quad(k=1,\dots,N_{\omega}),\quad
\Delta\omega=\frac{\omega_{\rm max}}{N_{\omega}-1},
\ee
\be
z_j=(j-1)\Delta z+z_{\rm min}\quad(j=1,\dots,N_z),\quad
\Delta z=\frac{z_{\rm max}-z_{\rm min}}{N_z-1}.
\ee
Here,
\be
N_{\omega}=100,\quad\omega_{\rm max}=0.99,\quad N_z=80,\quad
z_{\rm max}=40\,{\rm mm},\quad z_{\rm min}=0.5\,{\rm mm}.
\ee
For each $q$ we solve the linear inverse problem which is given by
\be
\rvv{y}(q)=M(q)\rvv{x}(q),
\ee
where $\rvv{y}(q)\in\Cm^{N_{\omega}}$, $\rvv{x}(q)\in\Cm^{N_z}$, $M(q)\in\Rm^{N_{\omega}\times N_z}$ are defined as
\be
\{\rvv{y}(q)\}_k=\Psi(q,\omega_k),\quad
\{\rvv{x}(q)\}_j=\widetilde{\eta}(q,z_j),
\ee
and
\be
\{M(q)\}_{kj}=
\frac{f_0\Delta z}{\ell^2}
\widetilde{G}_{\omega_k}(0,z_j,0)\widetilde{G}_{\omega_k}(q,0,z_j).
\ee

Each inverse problem is solved as (see Appendix \ref{pseudo})
\be
\rvv{x}(q)\approx M_{\rm reg}^+(q)\rvv{y}(q),
\ee
where $M_{\rm reg}^+(q)$ is the regularized pseudoinverse of $M(q)$. Let us discretize $q$ as
\be
q^{(k)}=\frac{2\pi}{h_d(2N_q+1)}k\quad(k=-N_q,\dots,N_q)
\ee
with
\be
N_q=16.
\ee
The truncated singular value decomposition was used to construct $M_{\rm reg}^+(q^{(k)})$ with the threshold $\sigma_0=10^{-4}$ (see Appendix \ref{pseudo}). As a result, 2 to 4 largest singular values were taken for each $q^{(k)}$.

Finally,
\be
\eta(x_i,z_j)=\frac{1}{h_d(2N_q+1)}\sum_{k=-N_q}^{N_q}e^{iq^{(k)}x_i}
\{\rvv{x}(q^{(k)})\}_j,\quad
i=1,\dots,N_d,\quad j=1,\dots,N_z.
\ee

\section{Results}

\begin{figure}[htbp]
\centering
\includegraphics[width=0.8\textwidth]{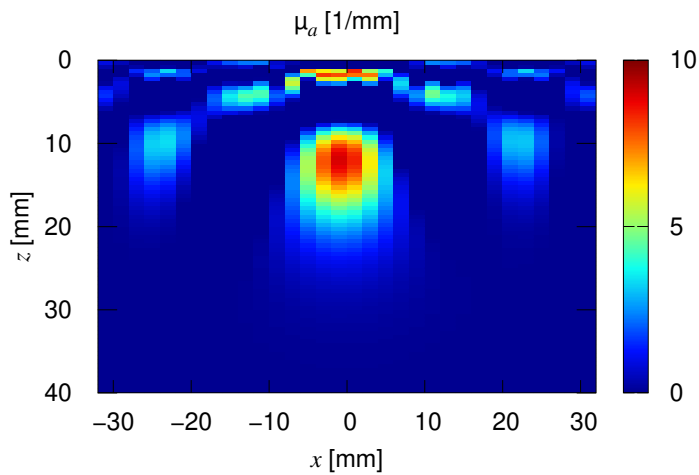}
\caption{
The reconstructed $\mu_a$ in (\ref{eqresult}). In the $x$-$z$ plane, true positions of four corners of the absorber rectangle are $(-6,4)$, $(-6,10)$, $(0,4)$, and $(0,10)$ in the unit of ${\rm mm}$.
}
\label{fig_recon}
\end{figure}

The reconstructed $\mu_a(\rvv{r})$ is obtained as
\begin{equation}
\mu_a(x_i,z_j)\approx\bar{\mu}_a+\eta(x_i,z_j).
\label{eqresult}
\end{equation}
The result is plotted in Fig.~\ref{fig_recon}.

\begin{figure}[htbp]
\centering
\includegraphics[width=0.45\textwidth]{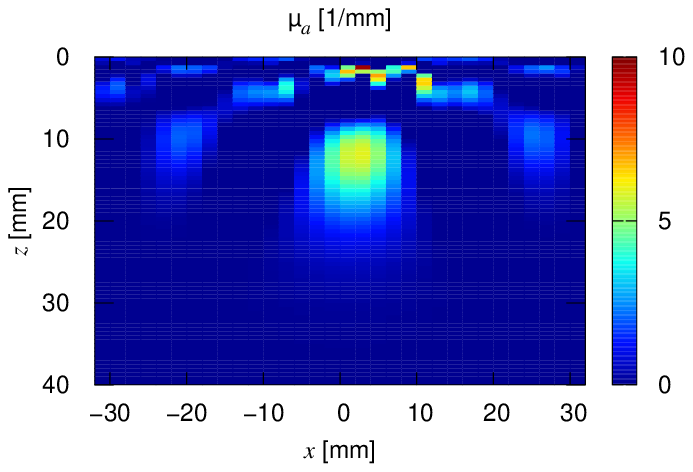}
\includegraphics[width=0.45\textwidth]{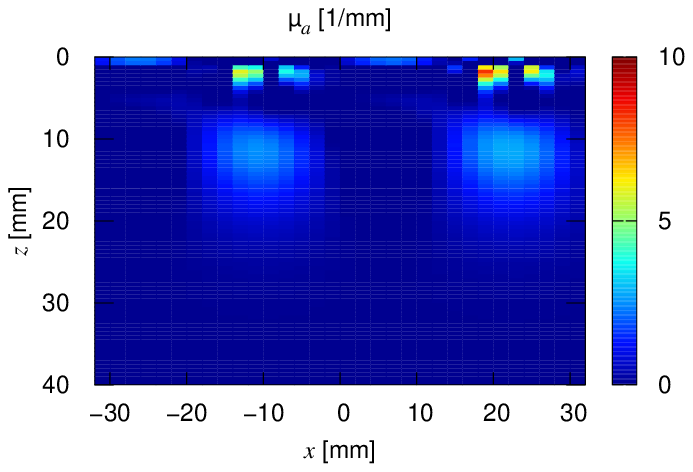}
\caption{
The reconstructed $\mu_a$ in (\ref{eqresult}) for (Left) $L=16\,{\rm mm}$ and (Right) $L=8\,{\rm mm}$. In the $x$-$z$ plane, true positions of four corners of the absorber rectangle are $(-6,4)$, $(-6,10)$, $(0,4)$, and $(0,10)$ in the unit of ${\rm mm}$.
}
\label{fig_recon2}
\end{figure}

In Fig.~\ref{fig_recon2}, the reconstruction was done for $L=16\,{\rm mm}$ ($N_f=8$, $\sigma_0=10^{-4}$) and $L=8\,{\rm mm}$ ($N_f=4$, $\sigma_0=10^{-2}$), respectively. The signal is weak and the reconstruction is vague when $L=16\,{\rm mm}$. For $L=8\,{\rm mm}$, the depth of the light propagation of the detected light is too shallow and the bar is not reconstructed. This behavior can be understood from the depth of the banana which was discussed in Secs.~\ref{banana} and \ref{psi}.

\section{Discussion}
\label{concl}

In this paper, the time-resolved measurement for the stripe illumination was proposed. The penetration depth of near-infrared light can be controlled with the pitch $L$ of the stripe. Although this fact is empirically well-known, we studied the situation in the first half of this paper by obtaining the depth of the banana shape. In \cite{Patterson-etal95}, $\langle z\rangle_{d_{\rm SD}}$ (see Sec.~2) was considered for a specific refractive-index ratio $\zeta=5.91$, which corresponds to $\mathfrak{n}\approx1.36$. In this paper, the banana depth $z_0$ is derived for general $\mathfrak{n}$. Although the formula (\ref{zsdformula}) was established from numerical experiments, $z_0$ in (\ref{z0value}) is derived from diffusion theory. We should emphasize that $z_0$ is introduced as the depth at which the point absorber most affects the detected light whereas the empirical relation $\langle z\rangle_{d_{\rm SD}}$ is the average depth. Hence, $z_0$ and $\langle z\rangle_{d_{\rm SD}}$ do not necessarily coincide even for $\mathfrak{n}=1.36$.

As shown in Sec.~\ref{banana}, a simple formula of $z_0=d_{\rm SD}/(2\sqrt{2})$ was obtained when the absorption is negligible for the zero boundary condition. The same formula was derived with a different way in the banana-shape regions theory \cite{Feng-etal95}. In this paper the formula $z_0$ is calculated for general cases.

In this paper, the diffuse optical tomography was tested by using the forward data from Monte Carlo simulation. The time-dependent measurement which is considered in this paper can be realized by single-photon avalanche diode (SPAD) arrays \cite{Bruschini-etal19}.

In Fig.~\ref{fig_recon}, the $x$ coordinate and size of the bar are reconstructed correctly. The $z$ coordinate of the bar is reconstructed at a position deeper than the actual depth of the bar. The reconstructed value of $\mu_a$ is overestimated. There are numerous reasons for the depth and value of the reconstructed absorption bar. The bright part of the stripe illumination was modeled by the Dirac delta function as shown in (\ref{afunc}), which does not have a width. The nonlinear inverse problem was linearized when tomographic images were computed. Indeed, the reconstruction was done in the transport regime, in which light propagation is governed by the radiative transport equation. Moreover, only the constant mode $m=0$ was taken into account for the operator $\widetilde{K}$ in (\ref{tildeK}).

Noisy reconstruction in Fig.~\ref{fig_recon} near the boundary is commonly known (for example, see \cite{Hebden-etal99}). A method to remove such high frequency noise near the boundary was proposed \cite{Srinivasan-etal04}.

Since the proposed optical tomography uses time-resolved data, a natural next step is the reconstruction of both the absorption and reduced scattering coefficients. The reconstruction of two parameters by the Rytov approximation \cite{Markel-Schotland04} will be utilized for the stripe illumination.

\section*{Acknowledgments}

This work was supported by JST, CREST Grant Number JPMJCR22C1, Japan. MM acknowledges support from JSPS KAKENHI Grant No.~JP18K03438 and JST PRESTO Grant Number JPMJPR2027. KK acknowledges support from JSPS KAKENHI Grant No.~JP16K04985, JP17H06102, JP18H01497, JP18H05240.

The Monte Carlo eXtreme (MCX) (\texttt{http://mcx.space/}) was used for Monte Carlo simulation. We truly thank Eiji Okada for providing Monte Carlo data for the banana-shape calculation in Sec.~\ref{banana}.

\section*{Declaration of interest statement}

The authors report there are no competing interests to declare.

\appendix

\section{Monte Carlo simulation}
\label{mcsimulation}

\begin{figure}[htbp]
\centering
\includegraphics[width=0.45\textwidth]{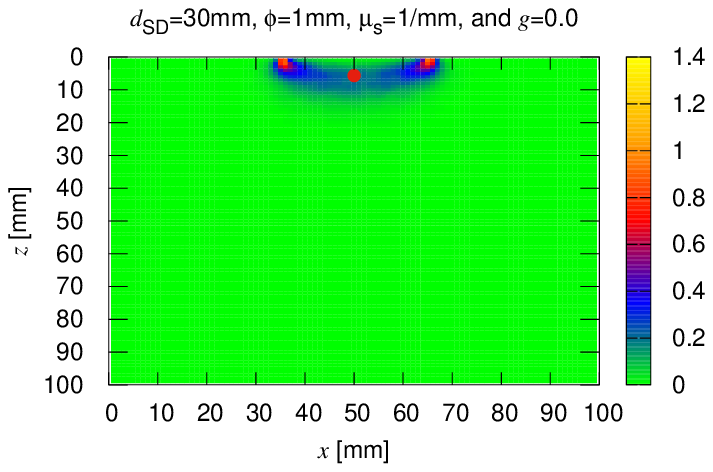}
\includegraphics[width=0.45\textwidth]{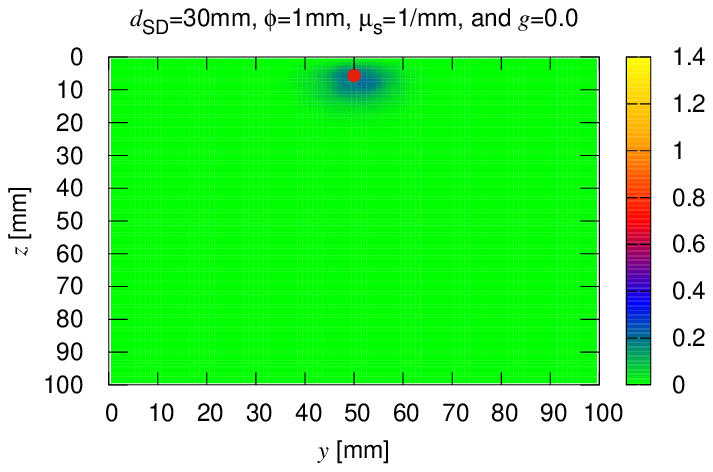}
\caption{
The spatial sensitivity profile is shown for $d_{\rm SD}=30\,{\rm mm}$, $\phi=1\,{\rm mm}$, $\mu_s=1\,{\rm mm}^{-1}$, and $g=0.0$. The cross sections at (Left) $y=50\,{\rm mm}$ and (Right) $x=50\,{\rm mm}$ are shown.
}
\label{mcfig3010}
\end{figure}

\begin{figure}[htbp]
\centering
\includegraphics[width=0.45\textwidth]{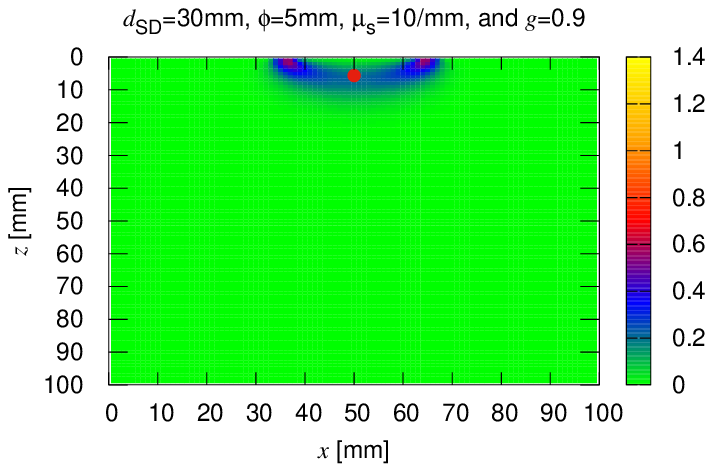}
\includegraphics[width=0.45\textwidth]{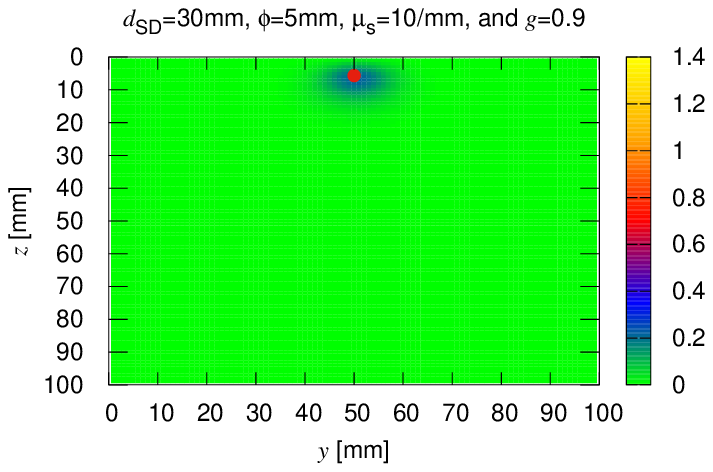}
\caption{
The spatial sensitivity profile is shown for $d_{\rm SD}=30\,{\rm mm}$, $\phi=5\,{\rm mm}$, $\mu_s=10\,{\rm mm}^{-1}$, and $g=0.9$. The cross sections at (Left) $y=50\,{\rm mm}$ and (Right) $x=50\,{\rm mm}$ are shown.
}
\label{mcfig3059}
\end{figure}

\begin{figure}[htbp]
\centering
\includegraphics[width=0.45\textwidth]{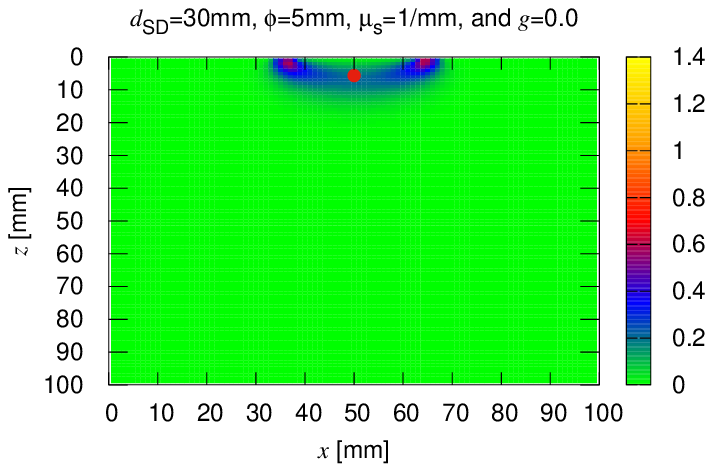}
\includegraphics[width=0.45\textwidth]{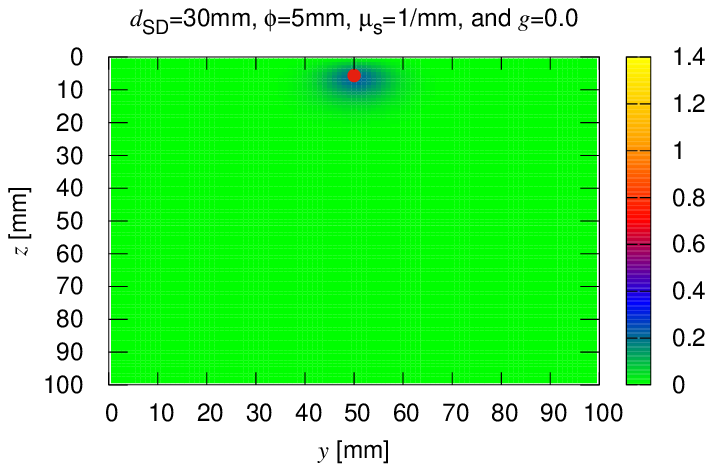}
\caption{
The spatial sensitivity profile is shown for $d_{\rm SD}=30\,{\rm mm}$, $\phi=5\,{\rm mm}$, $\mu_s=1\,{\rm mm}^{-1}$, and $g=0.0$. The cross sections at (Left) $y=50\,{\rm mm}$ and (Right) $x=50\,{\rm mm}$ are shown.
}
\label{mcfig3050}
\end{figure}

\begin{figure}[htbp]
\centering
\includegraphics[width=0.45\textwidth]{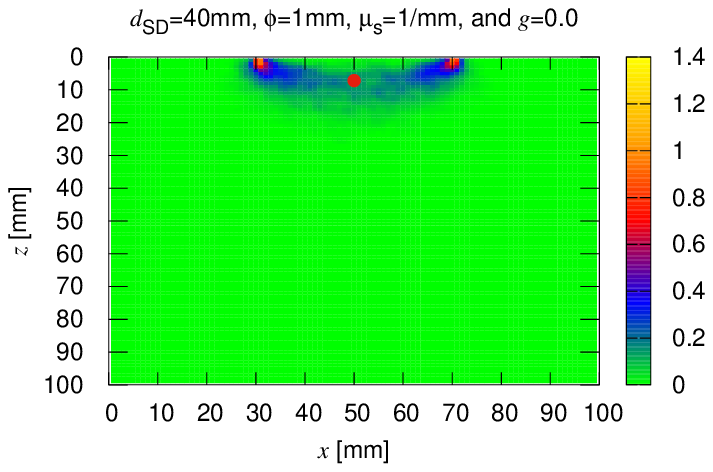}
\includegraphics[width=0.45\textwidth]{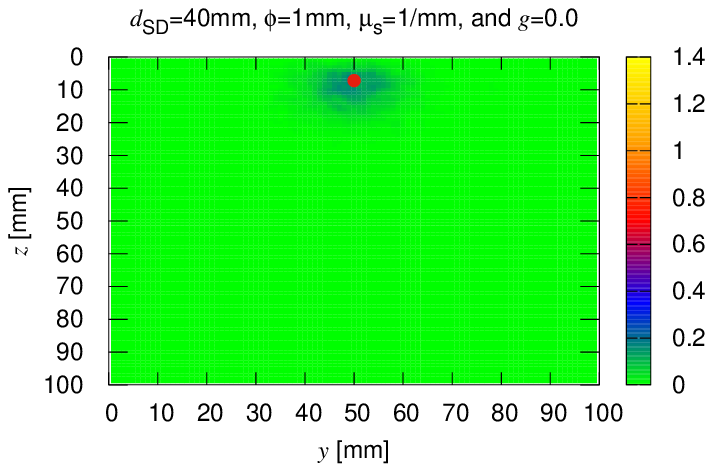}
\caption{
The spatial sensitivity profile is shown for $d_{\rm SD}=40\,{\rm mm}$, $\phi=1\,{\rm mm}$, $\mu_s=1\,{\rm mm}^{-1}$, and $g=0.0$. The cross sections at (Left) $y=50\,{\rm mm}$ and (Right) $x=50\,{\rm mm}$ are shown.
}
\label{mcfig4010}
\end{figure}

\begin{figure}[htbp]
\centering
\includegraphics[width=0.45\textwidth]{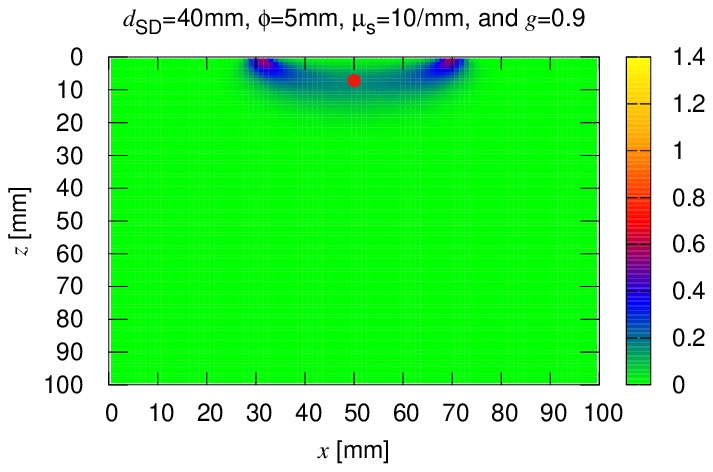}
\includegraphics[width=0.45\textwidth]{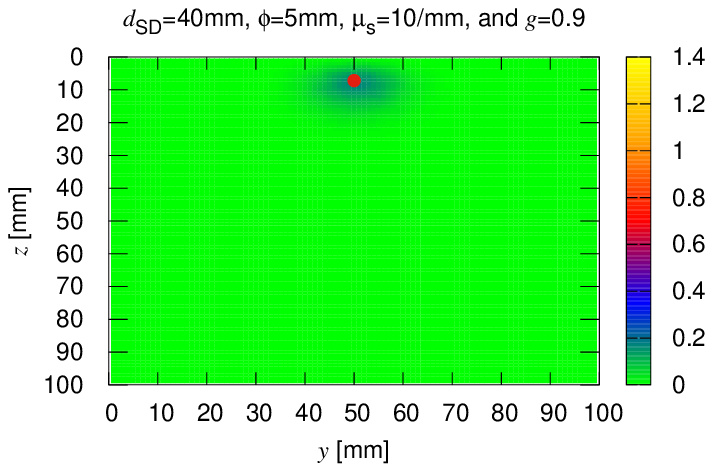}
\caption{
The spatial sensitivity profile is shown for $d_{\rm SD}=40\,{\rm mm}$, $\phi=5\,{\rm mm}$, $\mu_s=10\,{\rm mm}^{-1}$, and $g=0.9$. The cross sections at (Left) $y=50\,{\rm mm}$ and (Right) $x=50\,{\rm mm}$ are shown.
}
\label{mcfig4059}
\end{figure}

\begin{figure}[htbp]
\centering
\includegraphics[width=0.45\textwidth]{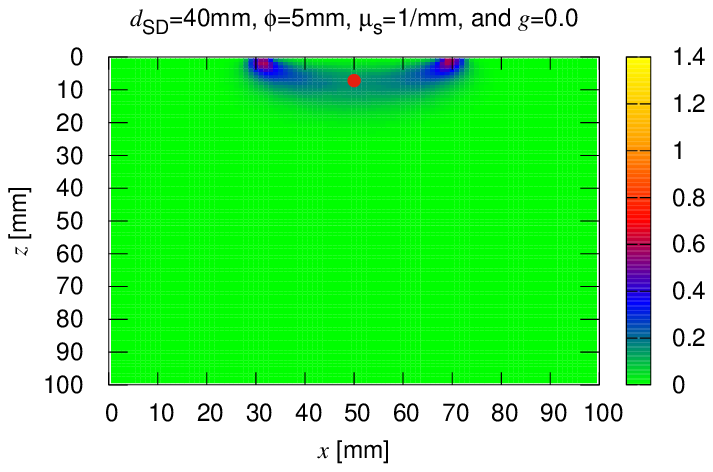}
\includegraphics[width=0.45\textwidth]{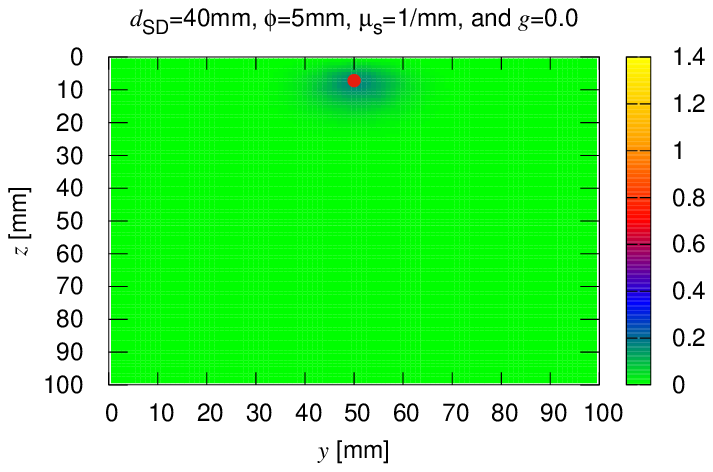}
\caption{
The spatial sensitivity profile is shown for $d_{\rm SD}=40\,{\rm mm}$, $\phi=5\,{\rm mm}$, $\mu_s=1\,{\rm mm}^{-1}$, and $g=0.0$. The cross sections at (Left) $y=50\,{\rm mm}$ and (Right) $x=50\,{\rm mm}$ are shown.
}
\label{mcfig4050}
\end{figure}

In Figs.~\ref{mcfig3010} through \ref{mcfig4050}, the position $z_0$ is compared with the spatial sensitivity profile from Monte Carlo simulation. The source-detector distance, diameter of optical fibers, scattering coefficient, and anisotropic factor are denoted by $d_{\rm SD}$, $\phi$, $\mu_s$, and $g$, respectively.

\section{Green's function}
\label{Agfunc}

Let us consider the Green's function which satisfies (\ref{Agfunc:def}). The Fourier transform is employed as
\begin{equation}
\widetilde{G}(\rvv{q},z,\rvv{r}')=\int_{\Rm^2}e^{-i\rvv{q}\cdot\rvv{\rho}}G(\rvv{r},\rvv{r}')\,d\rvv{\rho},
\label{fourierG}
\end{equation}
where $\rvv{q}\in\Rm^2$. We have
\be
\left\{\begin{aligned}
\pp_z^2\widetilde{G}-\lambda(q)^2\widetilde{G}=
\frac{-1}{D_0}e^{-i\rvv{q}\cdot\rvv{\rho}'}\delta(z-z'),
&\quad z>0,
\\
-z_e\pp_z\widetilde{G}+\widetilde{G}=0,
&\quad z=0.
\end{aligned}\right.
\ee

We can write
\be
\widetilde{G}=\left\{\begin{aligned}
A_1e^{\lambda z}+A_2e^{-\lambda z},&\quad 0<z<z',
\\
Be^{-\lambda z},&\quad z>z'.
\end{aligned}\right.
\ee
We have
\begin{equation}
(1-\lambda z_e)A_1+(1+\lambda z_e)A_2=0,
\label{cond1}
\end{equation}
\begin{equation}
A_1e^{\lambda z'}+A_2e^{-\lambda z'}=Be^{-\lambda z'},
\label{cond2}
\end{equation}
and the jump condition
\begin{equation}
-Be^{-\lambda z'}-A_1e^{\lambda z'}+A_2e^{-\lambda z'}=
\frac{-1}{\lambda D_0}e^{-i\rvv{q}\cdot\rvv{\rho}'}.
\label{cond3}
\end{equation}
From the conditions (\ref{cond1}), (\ref{cond2}), and (\ref{cond3}), we obtain
\ba
A_1&=
\frac{1}{2\lambda D_0}e^{-i\rvv{q}\cdot\rvv{\rho}'}e^{-\lambda z'},
\\
A_2&=
\frac{z_e\lambda-1}{z_e\lambda+1}
\frac{1}{2\lambda D_0}e^{-i\rvv{q}\cdot\rvv{\rho}'}e^{-\lambda z'},
\\
B&=
\frac{1}{2\lambda D_0}e^{-i\rvv{q}\cdot\rvv{\rho}'}
\left(e^{\lambda z'}+\frac{z_e\lambda-1}{z_e\lambda+1}e^{-\lambda z'}\right).
\ea

Let $z'\to0$. We obtain
\be
\widetilde{G}(\rvv{q},z,\rvv{r}')=
\frac{z_e}{D_0}e^{-i\rvv{q}\cdot\rvv{\rho}'}
\frac{1}{z_e\lambda(q)+1}e^{-\lambda(q)z},\quad z>0,\quad z'=0.
\ee
Therefore,
\ba
G(\rvv{r},\rvv{r}')
&=
\frac{1}{(2\pi)^2}\int_{\Rm^2}e^{i\rvv{q}\cdot\rvv{\rho}}
\widetilde{G}(\rvv{q},z,\rvv{r}')\,d\rvv{q}
\\
&=
\frac{z_e}{(2\pi)^2D_0}\int_{\Rm^2}e^{i\rvv{q}\cdot(\rvv{\rho}-\rvv{\rho}')}
\frac{e^{-\lambda(q)z}}{1+\lambda(q)z_e}\,d\rvv{q},\quad z>0,\quad z'=0.
\ea
We have
\ba
G(\rvv{r}_0,\rvv{r}_s)
&=
\frac{z_e}{(2\pi)^2D_0}\int_{\Rm^2}e^{-iq_x(x_0-x_s)}
\frac{e^{-\lambda(q)z_0}}{1+\lambda(q)z_e}\,d\rvv{q}
\\
&=
\frac{z_e}{(2\pi)^2D_0}\int_0^{2\pi}\int_0^{\infty}e^{-iq(x_0-x_s)\cos\va}
\frac{e^{-\lambda(q)z_0}}{1+\lambda(q)z_e}q\,dqd\va
\\
&=
\frac{z_e}{2\pi D_0}\int_0^{\infty}qJ_0(q|x_0-x_s|)
\frac{e^{-\lambda(q)z_0}}{1+\lambda(q)z_e}\,dq.
\ea
Similarly we have
\be
G(\rvv{r}_0,\rvv{r}_d)=
\frac{z_e}{2\pi D_0}\int_0^{\infty}qJ_0(q|x_0-x_d|)
\frac{e^{-\lambda(q)z_0}}{1+\lambda(q)z_e}\,dq.
\ee

\section{Computation of $H_{\omega}$}
\label{m2:doubleexp}

The function $H_{\omega}$ in (\ref{Homedef}) is written as
\be
H_{\omega}(x,z,z')=\int_0^{\infty}F(q;x,z,z')\,dq,
\ee
where
\be
F(q;x,z,z')
=\frac{1}{2\pi D_0Q_{\omega}(q)}\cos(qx)
\left[e^{-Q_{\omega}(q)|z-z'|}-\frac{1-Q_{\omega}(q)z_e}{1+Q_{\omega}(q)z_e}e^{-Q_{\omega}(q)|z+z'|}\right].
\ee
The integral can be evaluated by the double-exponential formula \cite{Ooura-Mori91,Ooura-Mori99,Ogata05}. Define
\be
\phi(\tau)=\frac{\tau}{1-e^{-6\sinh{\tau}}}
\ee
with
\be
\phi'(\tau)=\frac{1-(1+6\tau\cosh{\tau})e^{-6\sinh{\tau}}}{\left(1-e^{-6\sinh{\\
tau}}\right)^2}.
\ee
We have
\be
\int_0^{\infty}F(q;x,z,z')\,dq\approx
\frac{\pi}{|x|}\sum_{k=-N_k}^{N_k}F\left(\frac{\pi}{h|x|}\phi\left(kh+\frac{h}{2}\right);x,z,z'\right)
\phi'\left(kh+\frac{h}{2}\right),
\ee
where $N_k>0$ is an integer and $h$ is a mesh size.

\section{Pseudoinverse}
\label{pseudo}

Let us consider
\be
\rvv{\eta}=\underline{J}_{\rm reg}^+\rvv{\psi}.
\ee

\subsection*{C.1. Underdetermined}

In this case,
\be
\underline{J}_{\rm reg}^+=\underline{J}^*\underline{M}_{\rm reg}^{-1},
\quad \underline{M}=\underline{J}\,\underline{J}^*.
\ee
Here, $*$ denotes the Hermitian conjugate and ${\rm reg}$ means that the pesudoinverse is regularized by discarding singular values that are smaller than $\sigma_0$. Let $\sigma_j^2$ and $\rvv{v}_j$ be the eigenvalues and eigenvectors of the matrix $\underline{M}$:
\be
\underline{M}\rvv{z}_j=\sigma_j^2\rvv{z}_j.
\ee
We obtain
\be
\rvv{\eta}=\sum_{j\atop\sigma_j>\sigma_0}\frac{1}{\sigma_j^2}
\left(\rvv{z}_j^*\rvv{\psi}\right)\underline{J}^*\rvv{z}_j.
\ee

\subsection*{C.2. Overdetermined}

In this case,
\be
\underline{J}_{\rm reg}^+=\underline{M}_{\rm reg}^{-1}\underline{J}^*,
\quad \underline{M}=\underline{J}^*\,\underline{J}.
\ee
After solving the eigenproblem $\underline{M}\rvv{z}_j=\sigma_j^2\rvv{z}_j$, we obtain
\be
\rvv{\eta}_1=\sum_{j\atop\sigma_j>\sigma_0}\frac{1}{\sigma_j^2}
\left(\rvv{z}_j^*\underline{J}^*\rvv{\psi}\right)\rvv{z}_j.
\ee


\end{document}